\documentclass[11pt,english]{article}
\usepackage[utf8]{inputenc}
\usepackage{graphicx}
\usepackage{amsmath}
\usepackage{amssymb}
\usepackage{cite}
\usepackage{verbatim}
\usepackage{ragged2e,amsfonts,graphicx,caption,subcaption,xcolor}
\usepackage[left=2cm,right=2cm,top=3cm,bottom=3cm]{geometry}
\usepackage{hyperref}
\hypersetup{colorlinks=false, linkbordercolor={white}}
\makeatletter

\usepackage{babel}

\begin{document}
\begin{titlepage}
\thispagestyle{empty}

\bigskip

\begin{center}
\noindent{\Large \textbf {Thermodynamics of static and stationary black holes in Einstein-Gauss-Bonnet gravity  with dark matter
}}\\

\vspace{0,5cm}

\noindent{Í. D.D. Carvalho${}^{a}$\footnote{e-mail: icarodias@alu.ufc.br}, G. Alencar${}^{a}$\footnote{e-mail: geova@fisica.ufc.br} and C. R. Muniz${}^{b}$\footnote{e-mail: celio.muniz@uece.br} }

\vspace{0,5cm}

{\it ${}^a$Departamento de Física, Universidade Federal do Ceará-
Caixa Postal 6030, Campus do Pici, 60455-760, Fortaleza, Ceará, Brazil. \\
 }
 {\it ${}^b$\!Universidade Estadual do Cear\'a, Faculdade de Educa\c c\~ao, Ci\^encias e Letras de Iguatu, Iguatu-CE, Brazil.}
\end{center}

\vspace{0.3cm}

\begin{abstract}\noindent
This paper studies Einstein-Gauss-Bonnet (EGB) black holes surrounded by three phenomenological distributions of dark matter halos. The main result is obtaining the analytical solutions for the metric and all thermodynamic quantities, such as Hawking temperature, entropy, constant-volume heat capacity, and Gibbs free energy for static and stationary black hole solutions. Consequently, we determine a non-null horizon radius at which the black hole halts its evaporation by vanishing the temperature, indicating the emergence of remnants. In the stationary case, we also obtain the ergosphere regions for the found solutions and compare them. Finally, we find local and global phase transitions by studying the behavior of the heat capacity and Gibbs free energy.

\end{abstract}
\end{titlepage}

\tableofcontents
\newpage
\section{Introduction}\label{section-1}

In recent years, black hole physics research has received renewed interest. The measurement of gravitational waves by LIGO and VIRGO, the image of supermassive black holes by EHT, and the release of the James Webb Space Telescope brought worldwide attention to the subject\cite{LIGOScientific:2016aoc, EventHorizonTelescope:2022wkp, EventHorizonTelescope:2019dse, LIGOScientific:2017vwq}. The actual cosmological model for the universe, namely, $\Lambda$CDM, is well-supported by observational data, which have unveiled that the stuff of our universe consists of about $4\%$ baryonic matter, $29.6\%$ dark matter, and $67.4\%$ dark energy \cite{Planck:2018vyg}. Concerning the galaxies and their agglomerates, dark matter contributes to the formation, evolution, and coalescence of such structures through gravitational interaction \cite{Trujillo-Gomez:2010jbn}. However, it still is an open problem what is the true nature of that stuff (see \cite{Arbey:2021gdg} for a review). 

It is known that almost every large galaxy has a supermassive black hole in its center \cite{Kormendy:2013dxa}. It is worth mentioning that most of these are rotating black holes. Generally,  the galaxy matter is modeled by an anisotropic fluid with some density distribution. Dark matter dominates the almost spherical halo of these galaxies, with density distribution chosen according to the characteristics of the galaxy to be modeled \cite{BENSON201033}. In this direction, studies about black hole solutions surrounded by galactic halos were done, for example, ref. \cite{Konoplya:2022hbl} and their references. Dark matter halos are present in practically all galaxies, and thus one has considered the connection between the galactic central black holes and these halos (see, e.g., \cite{Zhang:2021bdr,Zhang:2022roh,Stuchlik:2022xtq}). Also, has been cogitated the possibility of the dark matter being the source for the black holes \cite{DasGupta:2015qbz,Xu:2018wow,Xu:2016ylr,Xu:2020jpv} and vice-versa, black holes as a source for dark matter, especially the primordial ones \cite{Frampton:2009yp,Carr:2021bzv,Mittal:2021egv,Villanueva-Domingo:2021spv,Escriva:2022duf}. Wormholes sourced by dark matter also has been considered in General Relativity, and modified gravity \cite{Sarkar:2019uhk,Jusufi:2019knb,Xu:2020wfm,Muniz:2022eex}. 

Modified Einstein--Gauss--Bonnet gravity has also drawn much attention recently. It is based on a generalization of Einstein's field equations by adding to the usual gravitational action higher derivatives terms in the form $S_{GB}=\int d^D x \sqrt{-g}\alpha\mathcal{G}$, where $\mathcal{G}$ is the so-called Gauss--Bonnet invariant term, given by \cite{Lovelock:1971yv}
\begin{equation}\label{GBLag}
\mathcal{G}=R^{\mu\nu}_{\rho\sigma}R_{\mu\nu}^{\rho\sigma}-4 R^{\mu}_{\nu}R^{\nu}_{\mu}+R^2,
\end{equation}
which represents a topological invariant term, {\it i.e.}, a total derivative in 3+1 dimensions, not yielding an effective gravitational dynamics. 

According to Lovelock's theorem \cite{Lovelock:1971yv, Lovelock:1972vz, Lanczos:1938sf}, Einstein's general relativity with the cosmological constant is the unique theory of gravity if we assume: (i) the space-time is $3+1$ dimensional, (ii) diffeomorphism invariance, (iii) metricity, and (iv) second order equations of motion. However, Glavan and Lin discovered a general covariant-modified theory of gravity in $D = 4$ space-time dimensions which propagates only the massless graviton and bypasses Lovelock's theorem. Their theory is formulated in $D >4$ dimensions. Its action consists of the Einstein-Hilbert term with a cosmological constant and the Gauss-Bonnet term multiplied by a factor $1/(D-4)$. The four-dimensional theory is defined as the limit $D \rightarrow 4$. In this singular limit, the Gauss-Bonnet invariant gives rise to non-trivial contributions to gravitational dynamics while preserving the number of graviton degrees of freedom and being free from Ostrogradsky instability \cite{Glavan:2019inb,Casalino:2020kbt}. A non-minimum coupling of the Einstein--Gaus--Bonnet Lagrangian with the scalar field is a consistent extension for such a theory and has been explored in some scenarios, including cosmological ones \cite{Odintsov:2021nim,Odintsov:2022zrj}. In this context, novel gravitational solutions were obtained and studied in several recent papers, as wormholes \cite{Jusufi:2020yus,Sharif:2020xxj,Canate:2022dzb,Barton:2022rkj}, black strings \cite{Suzuki:2022eaz}, black holes \cite{East:2021bqk,Gyulchev:2021dvt,Li:2021wqa,Rayimbaev:2022znx,Bravo-Gaete:2022mnr,Tsujikawa:2022lww, Kumar:2020xvu,Heydari-Fard:2020sib,Heydari-Fard:2021ljh}, gravitational collapse and the cosmic censorship conjuncture \cite{Yang:2020czk, Chatterjee:2021zre, Jaryal:2022rzd}, and black holes in connection with dark matter \cite{Fernandes:2021ysi}.

In this work, we study the thermodynamic properties of EGB gravitational theory considering dark matter profiles. We find two black hole solutions: a static and spherically symmetric one and a rotating one that depends on a generic dark matter density profile. We find analytically the thermodynamic quantities, temperature, entropy, heat capacity, and free energy for our two solutions considering three dark matter profiles: (i) NFW profile, (ii) Burkert profile, and (iii) pseudo-isothermal profile. Throughout this work, we consider $G = c = \hslash = k = 1$ and we will work in the convention $(-, +, +, +)$.

The paper is organized as follows. In section 2, we revisit the EGB theory and find the static and spherically symmetric solution of EGB black hole surrounded by an arbitrary distribution of matter. In section 3, we present the dark matter density profiles that we apply in our solution, then obtain the thermodynamic quantities and investigate their consequences. In section 4, we find the rotating black hole solutions, discuss their characteristics, and extend the thermodynamic analysis to these solutions. Finally, in section 5, we summarize the results. 

\section{Black hole solution in the EGB gravitational theory with a generic spherical distribution of matter}\label{section-2}
We can write the EGB action as 
\begin{eqnarray}\label{1}
\mathcal{S} = \frac{1}{16 \pi } \int d^{\text{D}}x \sqrt{-g} \left[R + \alpha \mathcal{G}\right],
\end{eqnarray}s
where $R$ is the Ricci scalar, $\mathcal{G}$ is given by Eq.(\ref{GBLag})  and $\alpha$ is the Gauss-Bonnet constant. 
Henceforth, we consider a \textit{ansatz} static and spherically symmetric spacetime in $4D$,
\begin{eqnarray}\label{3}
    ds^2 = -f(r)dt^2 + \frac{1}{f(r)}dr^2 +r^2\left(d\theta^2 + \sin^2{\theta}d\phi^2\right).
\end{eqnarray}
Although in $D = 4$, the GB term is a total derivative and hence does not contribute to the gravitational dynamics, an extra scalar field can be coupled with the GB term, known as Einstein-dilaton Gauss-Bonnet theory \cite{Kanti:1995vq, Kanti:1997br}, influencing that dynamic. However, recently Glavan and Lin \cite{Glavan:2019inb} studied the implications of GB term considering a rescaling of the coupling constant,
\begin{eqnarray}\label{4}
    \alpha \rightarrow \frac{\alpha}{D-4}.
\end{eqnarray}
In this case, the field equation of component $00$, taking into account a material source, becomes
\begin{eqnarray}\label{5}
    \frac{1}{r}\frac{df}{dr} + \frac{f}{r^2} - \frac{1}{r^2} - \alpha \left[\frac{2(f-1)}{r^3}\frac{df}{dr}-\frac{(f-1)^2}{r^4}\right] = -\rho(r),
\end{eqnarray}
where $\rho(r) = -T^0_0$. The exact solution of this equation is 
\begin{eqnarray}\label{6}
    f_{\pm}(r) = 1+\frac{r^2}{2\alpha}\left[1 \pm \sqrt{1 - \frac{4\alpha}{r^3}\left[const - \frac{E(r)}{4\pi}\right]}~\right],
\end{eqnarray}
where $E(r) = 4\pi\int dr r^2\rho(r)$. We want a solution that becomes the Schwarzschild solution when $\alpha \rightarrow 0$ and $\rho(r)=0$. So the solution which does that is $f_{-}(r)$ with $const = -2M$, explicitly
\begin{eqnarray}\label{7}
    f_{-}(r)  = 1+\frac{r^2}{2\alpha}\left[1 - \sqrt{1 + \frac{8M\alpha}{r^3} +\frac{\alpha E(r)}{4\pi r^3}}~\right],
\end{eqnarray}
with $\rho(r)$ i.e. $E(r)=0$, equation (\ref{7}) agrees with reference \cite{Guo:2020zmf}. The Einstein-Gauss-Bonnet field equations generate $T^0_0=T^1_1$ and $T^2_2=T^3_3$. The field equation for the component $22$ is 
\begin{eqnarray}\label{8}
    \frac{1}{2}\frac{d^2f}{dr^2}+ \frac{1}{r}\frac{df}{dr} - \frac{\alpha}{2}\left[\frac{2f}{r^2}\frac{d^2f}{dr^2}- \frac{2}{r^2}\frac{d^2f}{dr^2}+ \frac{2}{r^2}\left(\frac{df}{dr}\right)^2 - \frac{4(f-1)}{r^3}\frac{df}{dr} +\frac{2(f-1)^2}{r^4}\right] ={p_l},
\end{eqnarray}
where ${p_l} = T^2_2 = T^3_3$. After some manipulations and the use of equation (\ref{5}), we find 
\begin{eqnarray}\label{9}
    r\frac{d\rho}{dr} + 2\rho+2{p_l}=0.
\end{eqnarray}
Equation (\ref{9}) is the same as $\nabla_\mu T^\mu_r = 0$ and does not bring new physics. This way, the solution of equation (\ref{7}) is correct. 

\subsection{Thermodynamics of the EGB black hole solution surrounded by a generic spherical distribution of matter}\label{subsection2.1}
Now, we will investigate the thermodynamic properties of the solution of eq. (\ref{7}), from horizons given by the expression $f(r_h) = 0$. Then, the relationship between mass parameter $M$ and $r_h$ is
\begin{eqnarray}\label{16.2}
    M = \frac{r_h}{2}+\frac{\alpha}{2r_h} - \frac{E(r_h)}{32\pi}.
\end{eqnarray}
The Hawking temperature for the metric given by eq. (\ref{3}) is \cite{Wald:1984rg}
\begin{eqnarray}\label{17.2}
    T_H = \frac{1}{4\pi} \left(\frac{df}{dr}\right)_{r = r_h}.
\end{eqnarray}
Thus, we explicitly find the Hawking temperature as
\begin{eqnarray}\label{18.2}
    T_H = \frac{1}{4\pi r_h}\left[1 - \frac{\alpha}{r_h^2} - \frac{r_h^2}{4} \rho(r_h)~\right]\left({1 + \frac{2\alpha}{r_h^2}}\right)^{-1},
\end{eqnarray}
equation (\ref{18.2}) agrees with reference \cite{Guo:2020zmf} in limit of $\rho = 0$ and becomes the hawking temperature for Schwarzschild blackhole, $T_H = 1/4\pi r_h$, when $\alpha \rightarrow 0$ and $\rho = 0$.

Another quantity of interest is the entropy, which is obtained from $dS = dM/T_h$. We can compute it by using equations (\ref{16.2}) and (\ref{18.2}), finding
\begin{eqnarray}\label{19.2}
    dS = 2\pi r_h \left(1 + \frac{2\alpha}{r_h^2}\right) dr_h,
\end{eqnarray}
and then we can integrate it, arriving at
\begin{eqnarray}\label{20.2}
    S = \frac{A}{4} + 2\pi \alpha \ln\left(\frac{A}{A_0}\right),
\end{eqnarray}
where $A = 4\pi r_h^2$ is the horizon area and $A_0$ is a constant with dimension of area. An interesting inference is that the matter distribution surrounding the black hole does not influence the entropy, which is, therefore, the same as the EGB black hole vacuum solution \cite{Guo:2020zmf}. For more details on the discussions of the logarithmic behavior of the entropy, one can refer to \cite{Cai:2009ua}. 

Another quantity that brings physical insights is the heat capacity. We can compute this quantity as $C_v = dM/dT_H$, which results in
\begin{eqnarray}\label{21.2}
    C_v &=& 2\pi r_h^2 \left(1 + \frac{2\alpha}{r_h^2}\right)^2\left[1 - \frac{\alpha}{r_h^2} - \frac{r_h^2}{4} \rho(r_h)~\right]\times  \nonumber \\
    &\times& \left\{\frac{5\alpha}{r_h^2}-1+\frac{2\alpha^2}{r_h^4}-\frac{r_h^2}{4}\rho(r_h) \left[1+\frac{6\alpha}{r_h^2}\right]-\frac{r_h^3}{4}\frac{d\rho}{dr_h} \left[1+\frac{2\alpha}{r_h^2}\right]\right\}^{-1}
\end{eqnarray}

The last quantity we will compute is the Gibbs free energy $F = M - TS$. We find directly this quantity by using equations (\ref{16.2}), (\ref{18.2}) and (\ref{20.2}), finding
\begin{eqnarray}\label{22.2}
F =  \frac{r_h}{2}+\frac{\alpha}{2r_h} - \frac{E(r_h)}{32\pi} - \frac{1}{4\pi r_h}\left[1 - \frac{\alpha}{r_h^2} - \frac{r_h^2}{4} \rho(r_h)~\right]\left({1 + \frac{2\alpha}{r_h^2}}\right)^{-1} \left[\pi r_h^2 + 2\pi \alpha \ln\left(\frac{r_h^2}{r_0^2} \right)\right],
\end{eqnarray}
where $r_0$ is the radius in $A_0 = 4\pi r_0^2$. The equations (\ref{18.2}), (\ref{20.2}), (\ref{21.2}), (\ref{22.2}) cover, in an analytical way, all thermodynamic properties of an EGB black hole with a generic spherical distribution of matter surrounding it. The next subsection will analyze these thermodynamic properties of three dark matter density profiles.

\section{Thermodynamic properties of EGB black holes surrounded by dark matter}\label{section-3}
We are going to investigate the solution of equation (\ref{7}) and the thermodynamic properties discussed in subsection \ref{subsection2.1} concerning three dark matter profiles obtained from observations combined with numerical simulations.

\subsection{Dark matter density profiles}\label{subsection3.1}
The first one is the Burkert density profile \cite{Burkert:1995yz}. This is a cored profile of dark matter halos, which is popular due to its agreement with phenomenology \cite{Gentile:2004tb,Gentile:2006hv,Salucci:2007tm,Rodrigues:2017vto}. Its density is 
\begin{eqnarray}\label{10}
    \rho_B(r) = \frac{\rho_c}{\left(1 + \frac{r}{r_c}\right)\left(1 + \frac{r^2}{r_c^2}\right)}.
\end{eqnarray}
The Burkert density profile is approximately constant for small radial distances from the center and is proportional to $r^{-3}$ to the large ones. The $E_B = 4\pi \int dr r^2 \rho_B(r)$, explicitly
\begin{eqnarray}\label{11}
    E_B(r) = \pi \rho_c r_c^3\left[2 \ln\left(1+\frac{r}{r_c}\right) + \ln\left(1+\frac{r^2}{r_c^2}\right) - 2 \arctan\left(\frac{r}{r_c}\right)\right].
\end{eqnarray}
The second dark matter density profile is due to Navarro,  Frenk, and White (NFW); they found this profile by studying the equilibrium density of dark matter halos using high-resolution N-body simulations. They found that all profiles should have the same shape regardless of the halo mass, initial density fluctuation spectrum, and the values of the cosmological parameters \cite{Navarro:1995iw,Navarro:1996gj}. The dark matter profile density is 
\begin{eqnarray}\label{12}
    \rho_{NFW}(r) = \frac{\rho_s}{\frac{r}{r_s}\left(1 + \frac{r}{r_s}\right)^2},
\end{eqnarray}
where $\rho_s$ and $r_s$ are two characteristic parameters of the model. Note that for large radial distances from the center, the NFW density profile is proportional to $r^{-3}$ like the Burkert profile. The NFW profile increases without limit when this distance is small, but the mass $E_{NFW}(r)$ is finite for all $r$. It is given by 
\begin{eqnarray}\label{13}
    E_{NFW}(r) = \pi \rho_s r_s^3\left[\ln\left(1+\frac{r}{r_s}\right) + \frac{1}{1+\frac{r}{r_s}}\right].
\end{eqnarray}
The last dark matter profile we will analyze is the pseudo-isothermal profile \cite{Begeman:1991iy}. The corresponding dark matter profile density is 
\begin{eqnarray}\label{14}
    \rho_{iso}(r) =\rho_c \left[1 + \frac{r^2}{r_c^2}\right]^{-1}.
\end{eqnarray}
This profile does not behave similarly to the two others to large radial distances. Namely, its behavior in this limit is proportional to $r^{-1}$. On the other hand, when $r$ is small, this profile approximates to $\rho_c$, i.e., the core density. The integrated mass is 
\begin{eqnarray}\label{15}
E_{iso}(r) = 4\pi \rho_c r_c^3 \left[\frac{r}{r_c} - \arctan\left(\frac{r}{r_c}\right)\right].
\end{eqnarray}
It is worth pointing out that, in the limit for large distances, such a mass will cause Keplerian circular trajectories with approximately constant orbital velocities, like the ones observed in the curves of galaxy rotations.

\begin{figure}[!h]
    \centering
    \includegraphics[width=10.0cm]{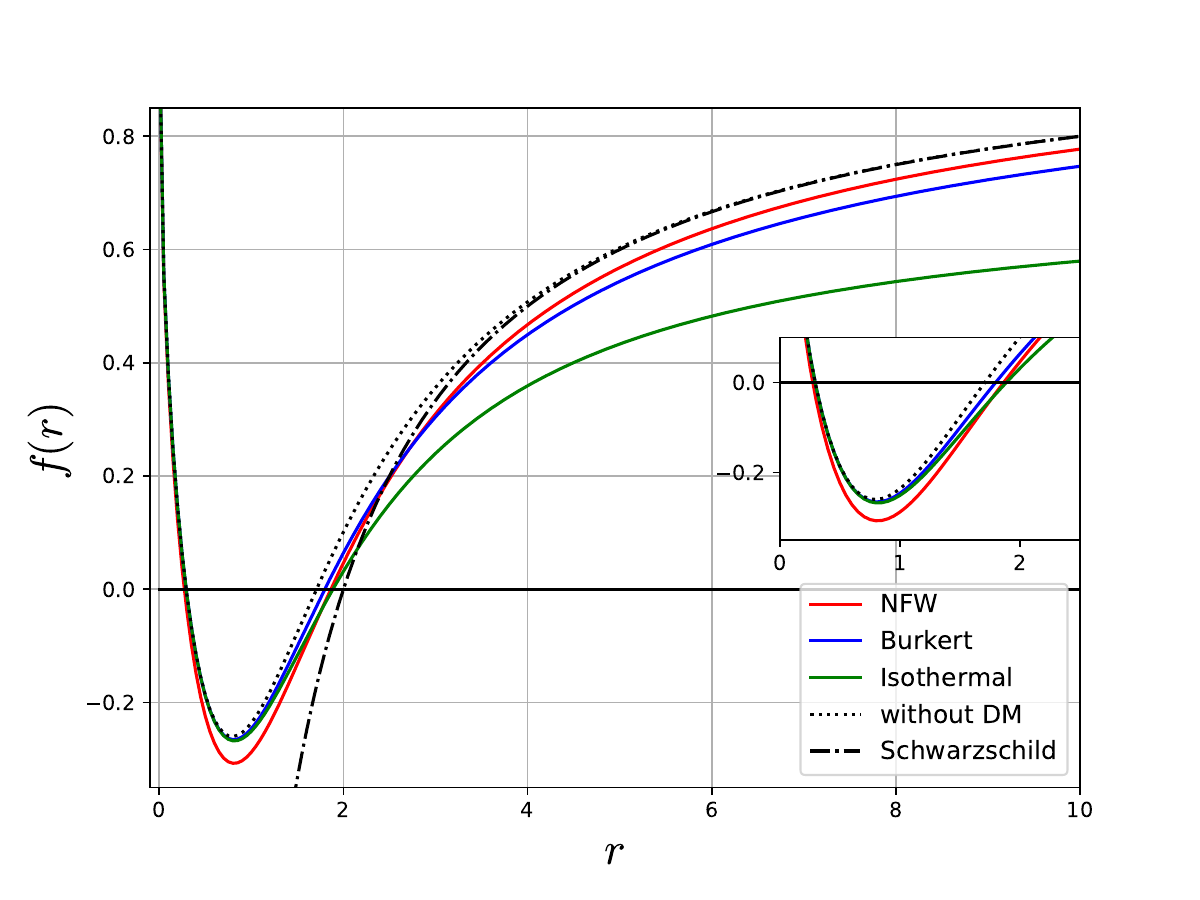}
    \caption{$f(r)$ for differents DM profiles. $\rho_c=\rho_s=0.5$, $r_c=r_s=1.5$, $M=1$ and $\alpha =0.5$ }
    \label{figure1}
\end{figure}
\subsection{Thermodynamic properties of EGB black holes with dark matter}\label{subsection3.2}

\begin{figure}[!h]
    \centering
    \includegraphics[width=10.0cm]{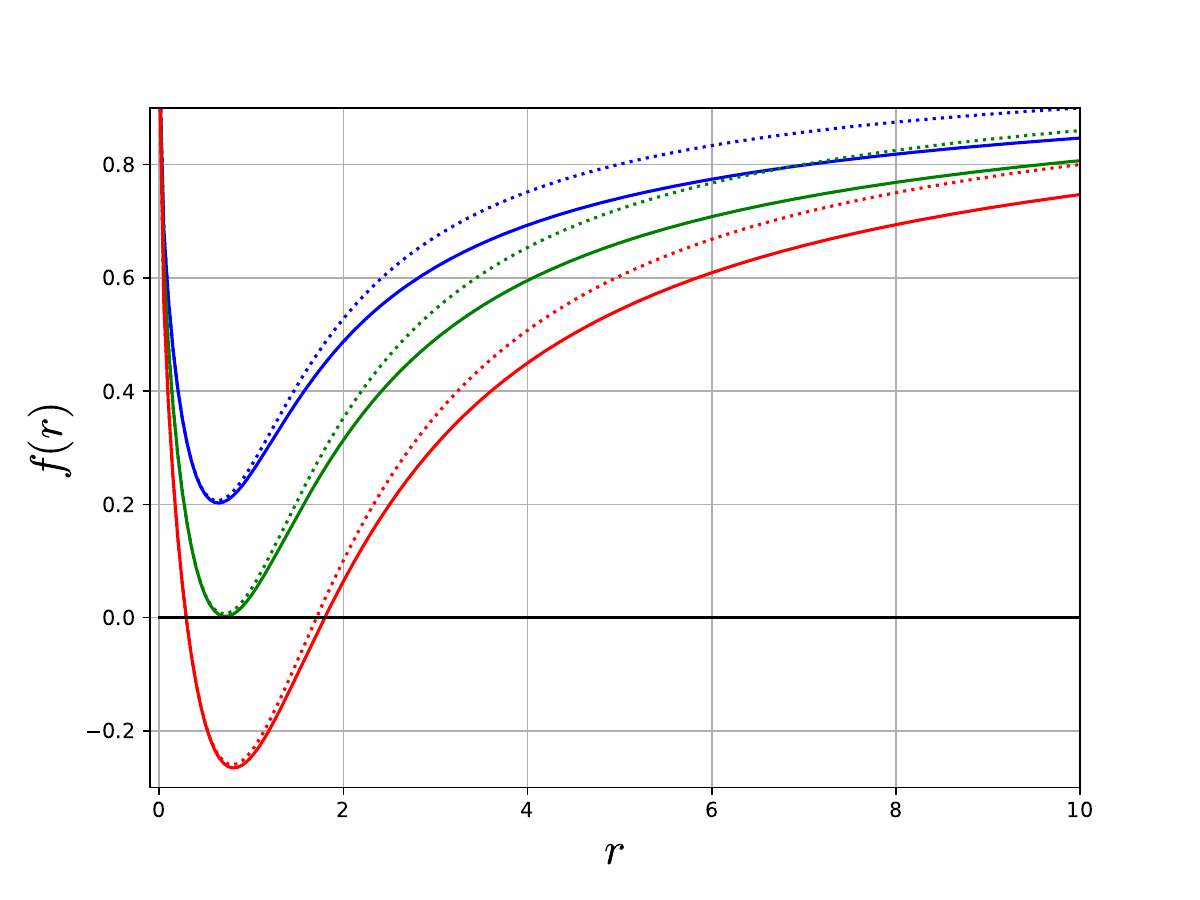}
    \caption{$f(r)$ for differents DM profiles. $\rho_c=\rho_s=0.5$, $r_c=r_s=1.5$ and $\alpha =0.5$. The solid line is the metric function with Burkert dark matter, and the dotted line is the metric function without dark matter. The blue ones has $M = 0.5$, the green ones has $M = 0.7$, and the red ones has $M = 1$ }
    \label{figure1.2}
\end{figure}

\begin{figure}[!h]
    \centering
    \includegraphics[width=10cm]{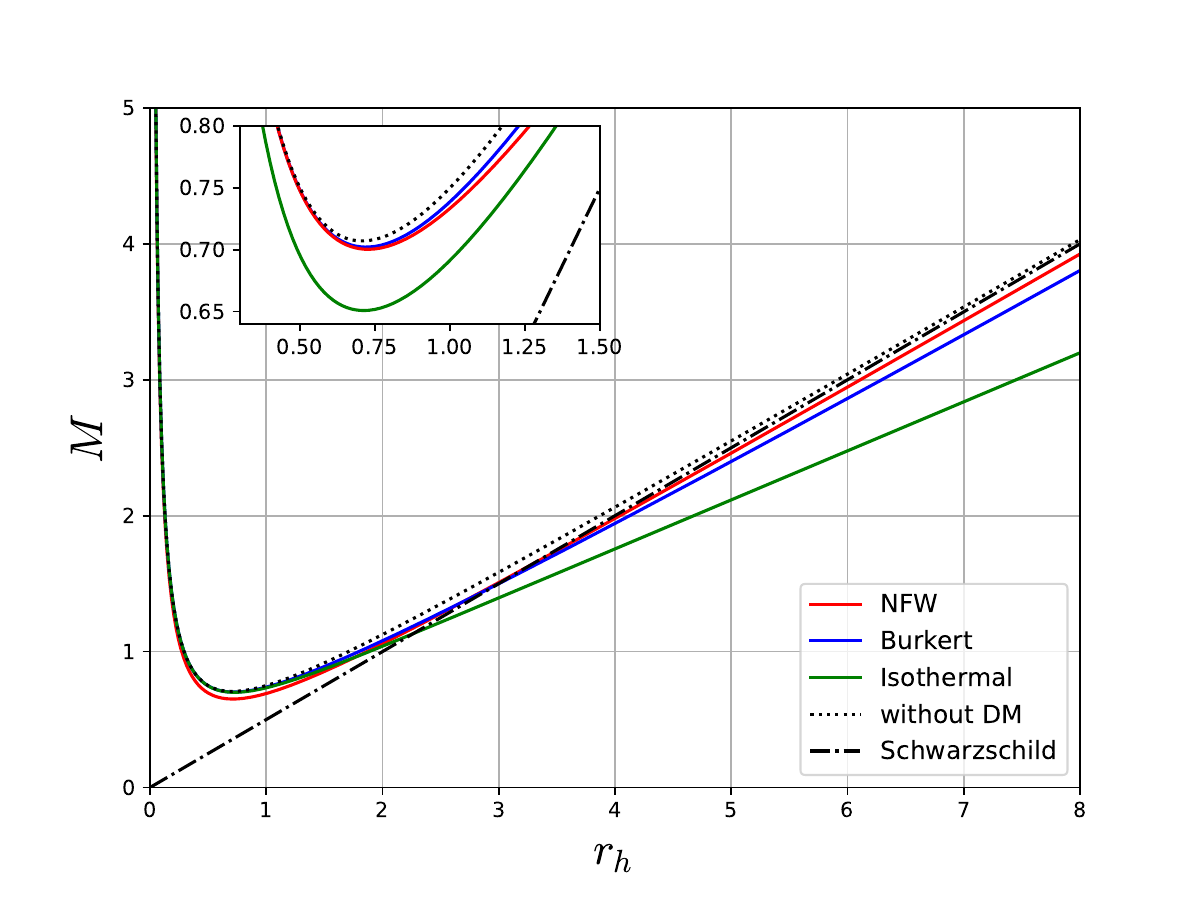}
    \caption{Behavior of $M$ for differents DM profiles, $\rho_c=\rho_s=0.5$, $r_c=r_s=1.5$ and $\alpha =0.5$.}
    \label{figure2}
\end{figure}

Now, we will investigate the thermodynamic properties of the EGB black hole to different dark matter profiles presented in section \ref{subsection3.1}. In figure \ref{figure1}, we can see the behavior of $f(r)$ for each dark matter profile. To appropriate values of the parameters, there are horizons given by $f(r_h) = 0$, and according to these parameters, it is possible to obtain two, one, or no horizon. We can also see the relationship between the mass parameter $M$ and the horizon radius through equation (\ref{16.2}) with $E(r)= 4\pi\int dr r^2 \rho_{DM}(r)$, where $\rho_{DM}(r)$ is one of the dark matter density profiles, see figure \ref{figure2}. Solving equation (\ref{16.2}) to find all horizons' radii, in case of their existence, can be very hard due to the dependence of dark matter manifested in $E(r)$. Then we made figure \ref{figure1.2} to exemplify the possibility of none, one, or two horizons' radii. In figure \ref{figure1.2}, we plot the set of metric functions with the Burkert profile, and without dark matter, each set has a different value of the mass parameter: $M = 0.5$, $ M = 0.7$, $M=1$. The behavior of the other dark matter profiles is similar.

\begin{figure}[!h]
    \centering
    \includegraphics[width=10.0cm]{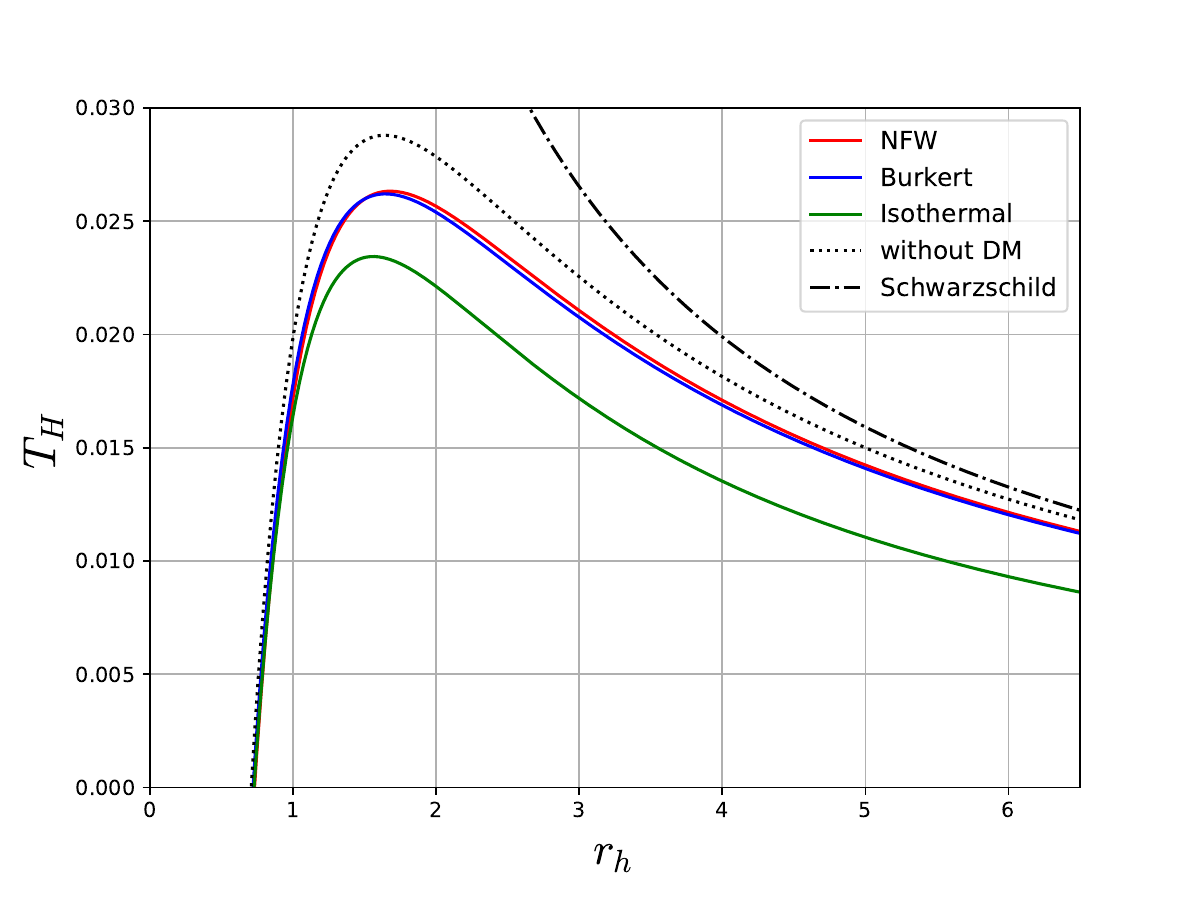}
    \caption{Behavior of $T_H$ for differents DM profiles, $\rho_c=\rho_s=0.5$, $r_c=r_s=1.5$ and $\alpha =0.5$.}
    \label{figure3}
\end{figure}

Let us investigate the Hawking temperature for the dark matter profiles. As already discussed, the Hawking temperature for a generic spherical distribution of matter is given by equation (\ref{18.2}). We can see the behavior of the temperature to three dark matter profiles in figure \ref{figure3}. The temperature is equal to zero for the horizon radius which satisfies
\begin{eqnarray}\label{16}
    4r_h^2 - 4\alpha - r_h^4 \rho(r_h) = 0,   
\end{eqnarray}
in a specific way, the temperature vanishes for each dark matter profile for all $r_h$ which satisfies the equation (\ref{16}), where $\rho(r)$ is the dark matter profile density. When the black hole reaches this critical $r_h$, the evaporation ceases, and a remnant survives. For instance, on considering the pseudo-isothermal dark matter profile (\ref{14}), this critical horizon radius is given by
\begin{equation}\label{16.5}
   r_h^c= \sqrt{2} \sqrt{\frac{\sqrt{\alpha^2+\alpha r_c^4 \rho_c +2 \alpha r_c^2+r_c^4}+\alpha-r_c^2}{r_c^2 \rho_c +4}}.
\end{equation}
Notice that, in this case, for $\alpha=0$ ({\it i.e.}, in general relativity), there is no such remnant, since $r_h^c=0$, even in the presence of dark matter.

As already stated in subsection \ref{subsection2.1}, the entropy is not changed by a generic spherical distribution of matter surrounding the EGB black hole. Thus, the entropy for the three dark matter profiles is the same as equation (\ref{20.2}).

\begin{figure}[!h]
    \centering
    \includegraphics[width=10.0cm]{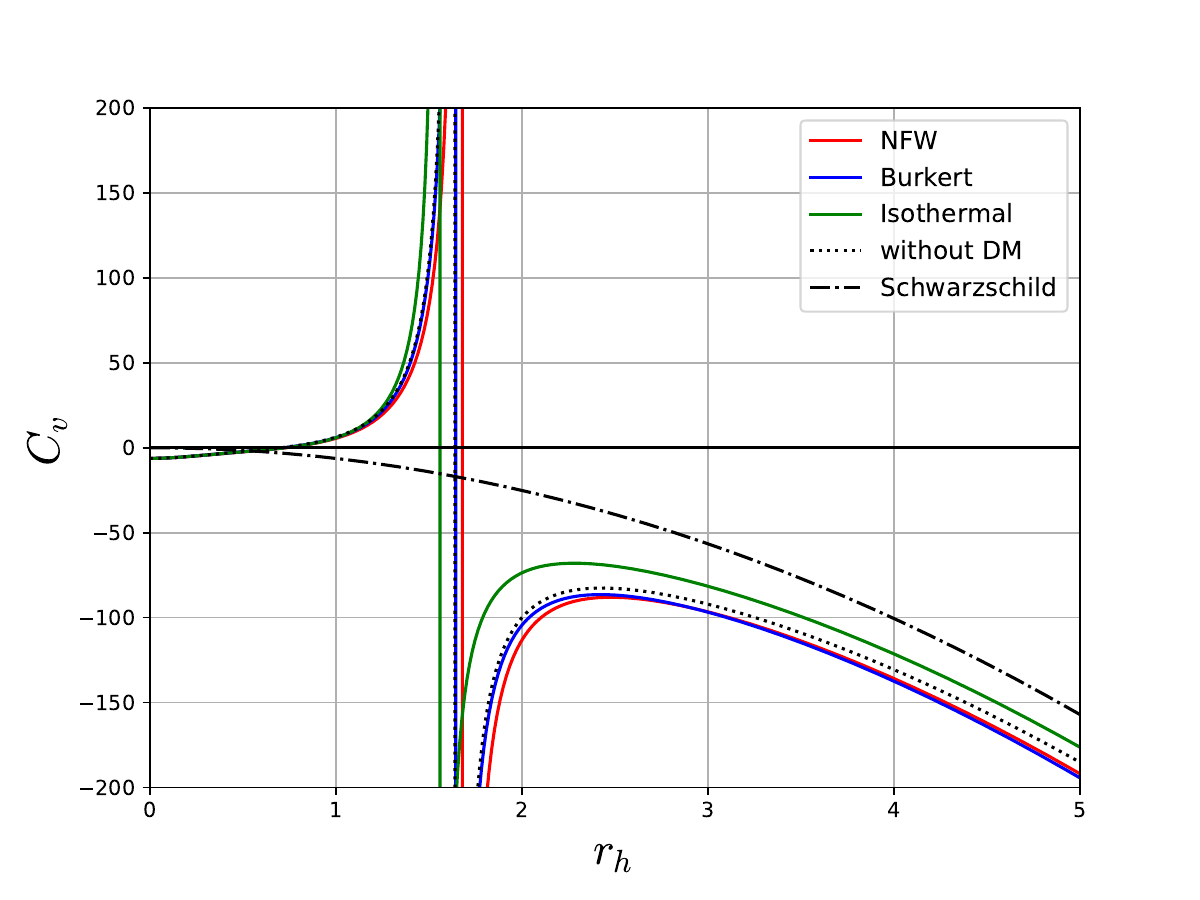}
    \caption{Behavior of $C_v$ for different DM profiles, $\rho_c=\rho_s=0.5$, $r_c=r_s=1.5$ and $\alpha =0.5$.}
    \label{figure4}
\end{figure}

Let us now present the heat capacity results at constant volume for the three dark matter profiles. The behavior of this quantity can be seen in figure \ref{figure4}. The black hole presents local phase transitions when the heat capacity vanishes, {\it i.e.}, at the horizon radius which obeys 
\begin{eqnarray}\label{17}
    \frac{5\alpha}{r_h^2}-1+\frac{2\alpha^2}{r_h^4}-\frac{r_h^2}{4}\rho(r_h) \left[1+\frac{6\alpha}{r_h^2}\right]-\frac{r_h^3}{4}\frac{d\rho}{dr_h} \left[1+\frac{2\alpha}{r_h^2}\right] = 0,
\end{eqnarray}
where $\rho$ is the dark matter profile density.
Notice that the black holes present smooth (abrupt) phase transitions from unstable (stable) regions to stable (unstable) ones ($C_v<0$, unstable; $C_v>0$, stable).

\begin{figure}[!h]
    \centering
    \includegraphics[width=10.0cm]{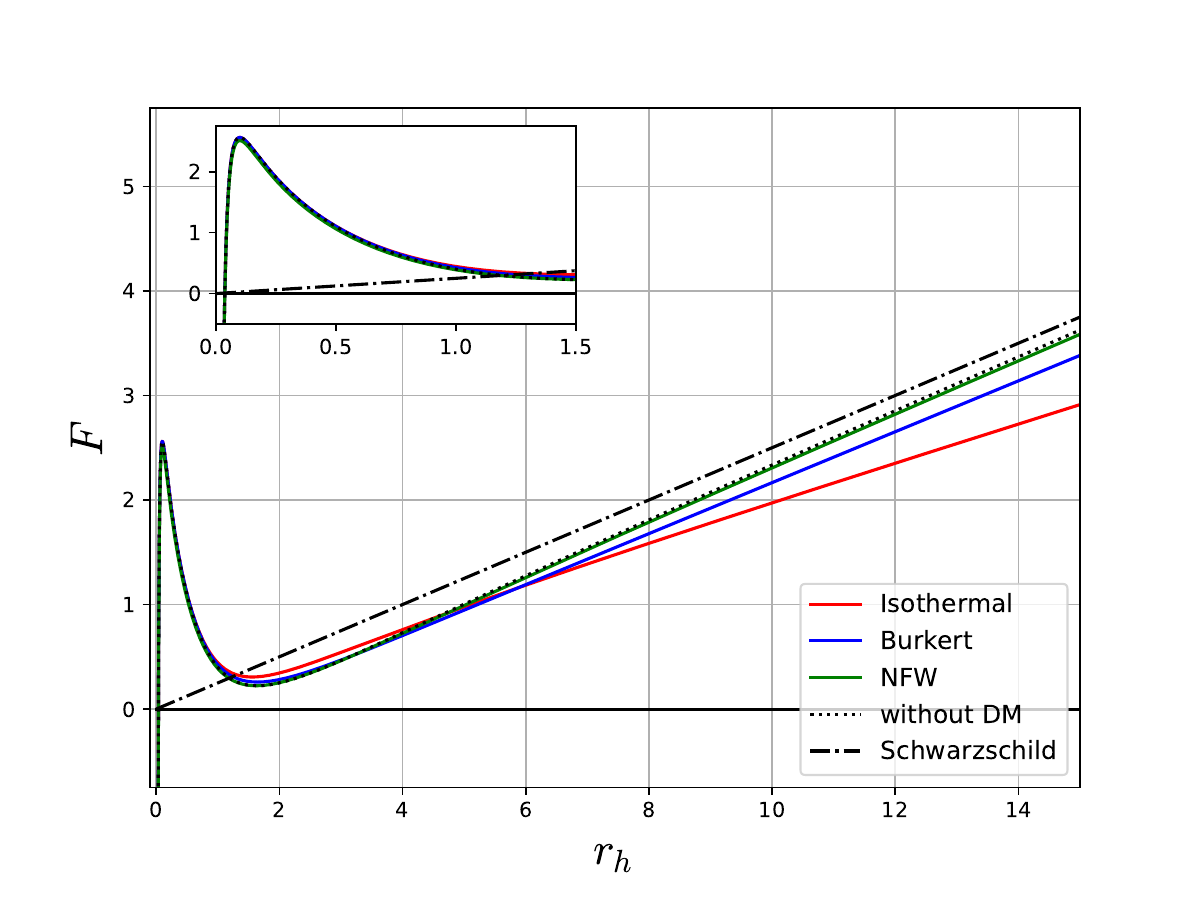}
    \caption{Behavior of $F$ for different DM profiles, $\rho_c=\rho_s=0.5$, $r_c=r_s=1.5$, $r_0=0.1$ and $\alpha =0.5$}
    \label{figure5}
\end{figure}

We also consider the Gibbs free energy, equation (\ref{22.2}), for each dark matter profile. The behavior of this quantity can be seen in figure \ref{figure5}.
We can see from the plot that the Gibbs free energy encompasses only smooth (global) phase transitions ($F<0$, stable; $F>0$, unstable).

\section{Thermodynamic properties of EGB rotating black holes surrounded by dark matter}
\label{rotating}
As the introduction states, most of the supermassive black holes in the center of galaxies are rotating. Therefore, this is the more relevant case, and we analyze it here. 
\subsection{Newman-Janis algorithm}
To obtain the rotating counterpart of the black hole surrounded by dark matter in 4D EGB gravity (BHDM-EGB), we use the seed metric (\ref{3}) and apply the Newman-Janis algorithm, following \cite{Papnoi:2021rvw}. Initially, let us introduce the metric (\ref{3}) in the Eddington-Finkelstein coordinates ($u$, $r$, $\theta$, $\phi$) with

\begin{equation}\label{tr}
 du=dt-\frac{dr}{f(r)}.
\end{equation}
From the transformation (\ref{tr}), the metric of the non-rotating BHDM-EGB becomes
\begin{equation}
 ds^2=f(r)du^2+2dudr-r^2d\theta^2-r^2\sin^2\theta d\phi^2.\label{sur}
\end{equation}
Next, the metric can be expressed in terms of null tetrad as
\begin{equation}
 g^{ab}=l^am^b+l^bn^a-m^a\bar{m}^b-m^b\bar{m}^a.
\end{equation}
Notice that $l^a$ and $n^a$ are real, $m^a$, $\bar{m^a}$ are mutual complex conjugate. Now, the null tetrad of the metric must be in the form 
\begin{eqnarray}
 l^a&=&\delta^a_r,\\
 n^a&=&\delta^a_\mu-\frac{f(r)}{2}\delta^a_r,\\
 m^a&=&\frac{1}{\sqrt{2}r}\left(\delta^a_\theta+\frac{i}{\sin\theta}\delta^a_\phi\right).
\end{eqnarray}
The null tetrad obeys null, orthogonal, and metric
conditions as follows
\begin{eqnarray}
 l^al_a=n^an_a=m^am_a=\bar{m}^a\bar{m}_a=0,\\
 l^am_a=l^a\bar{m}_a=n^am_a=n^a\bar{m}_a=0,\\
 l^an_a=-m^a\bar{m}_a=1.
\end{eqnarray}
Now, we perform the complex coordinate transformations in ($u$, $r$)-plane by using the relations
\begin{eqnarray}
 && u'\rightarrow u-ia\cos\theta,\nonumber\\
 && r'\rightarrow r+ia\cos\theta,\label{rp}
\end{eqnarray}
with $a$ being the rotating parameter of the BH. The complexification of the radial coordinate results in the change in the metric functions of (\ref{sur}) to new undetermined ones. Now, we make the transformation
\begin{eqnarray}
 &&f(r)\rightarrow F(r, a, \theta),\\
 &&r^2\rightarrow H(r, a, \theta),
\end{eqnarray}
 and thus, the null tetrad with the rotation parameter $a$ becomes
\begin{eqnarray}\label{nt}
 l^a&=&\delta^a_r,\\
 n^a&=&\delta^a_\mu-\frac{F}{2}\delta^a_r,\\
 m^a&=&\frac{1}{\sqrt{2H}}\left((\delta^a_\mu-\delta^a_r)ia\sin\theta+\delta^a_\theta
 +\frac{i}{\sin\theta}\delta^a_\phi\right).
\end{eqnarray}
Thus, applying the new null tetrad (\ref{nt}), the stationary metric for BHDM-EGB in the Eddington-Finkelstein coordinates obtained is
\begin{eqnarray}\label{mt}
 ds^2&=& Fdu^2+2du dr+2a\sin^2\theta(1-F)du d\phi \nonumber \\ && -2a\sin^2\theta dr d\phi -H d\theta^2 \nonumber\\
 && -\sin^2\theta\left(H+a^2\sin^2\theta(1-F)\right)d\phi^2.
\end{eqnarray}
Now we change the metric (\ref{mt}) into the Boyer-Lindquist coordinates. We obtain the rotating EGB BH by introducing the transformation
\begin{eqnarray}
 du&=&dt+\nu(r)dr,\\
 d\phi&=&d\phi'+\chi(r)dr,
\end{eqnarray}
with 
\begin{eqnarray}
 \nu(r)&=&-\frac{a^2+r^2}{a^2+r^2f(r)},\\
 \chi(r)&=&-\frac{a}{a^2+r^2f(r)}.
\end{eqnarray}
By choosing
\begin{eqnarray}
 F&=&\frac{(r^2f(r)+a^2\cos^2\theta)}{H},\;\;\;
 H=r^2+a^2\cos^2\theta.
\end{eqnarray}
Hence, the stationary 4D BHDM-EGB metric reads as
\begin{eqnarray}
 ds^2 &=& -\frac{\Delta}{\Sigma}(dt-a\sin^2\theta d\phi)^2+\frac{\Sigma}{\Delta}dr^2+\Sigma d\theta^2
  \nonumber \\ && +\frac{\sin^2\theta}{\Sigma}\left(a dt-(r^2+a^2)d\phi\right)^2.\label{Romet}
\end{eqnarray}
For the metric (\ref{Romet}), the metric functions are
\begin{eqnarray}
 \Sigma&=&r^2+a^2\cos^2\theta, \label{SIGMA}\\
 \Delta&=&r^2+a^2+\frac{r^4}{2\alpha}\left[1-\sqrt{1+\frac{8M\alpha}{r^3}+\frac{\alpha E(r)}{4\pi r^3}}\right],\label{DDm}
\end{eqnarray}
where $E(r)=\int_0^r 4\pi r^2\rho_{dm}dr$. Therefore, each dark matter profile yields a different rotating black hole. Notice that in the absence of dark matter, $E(r)=0$, we retrieve the 4D EGB rotating black hole solution obtained in \cite{Papnoi:2021rvw}.

\begin{figure}[!h]
    \centering
    \includegraphics[width=10.0cm]{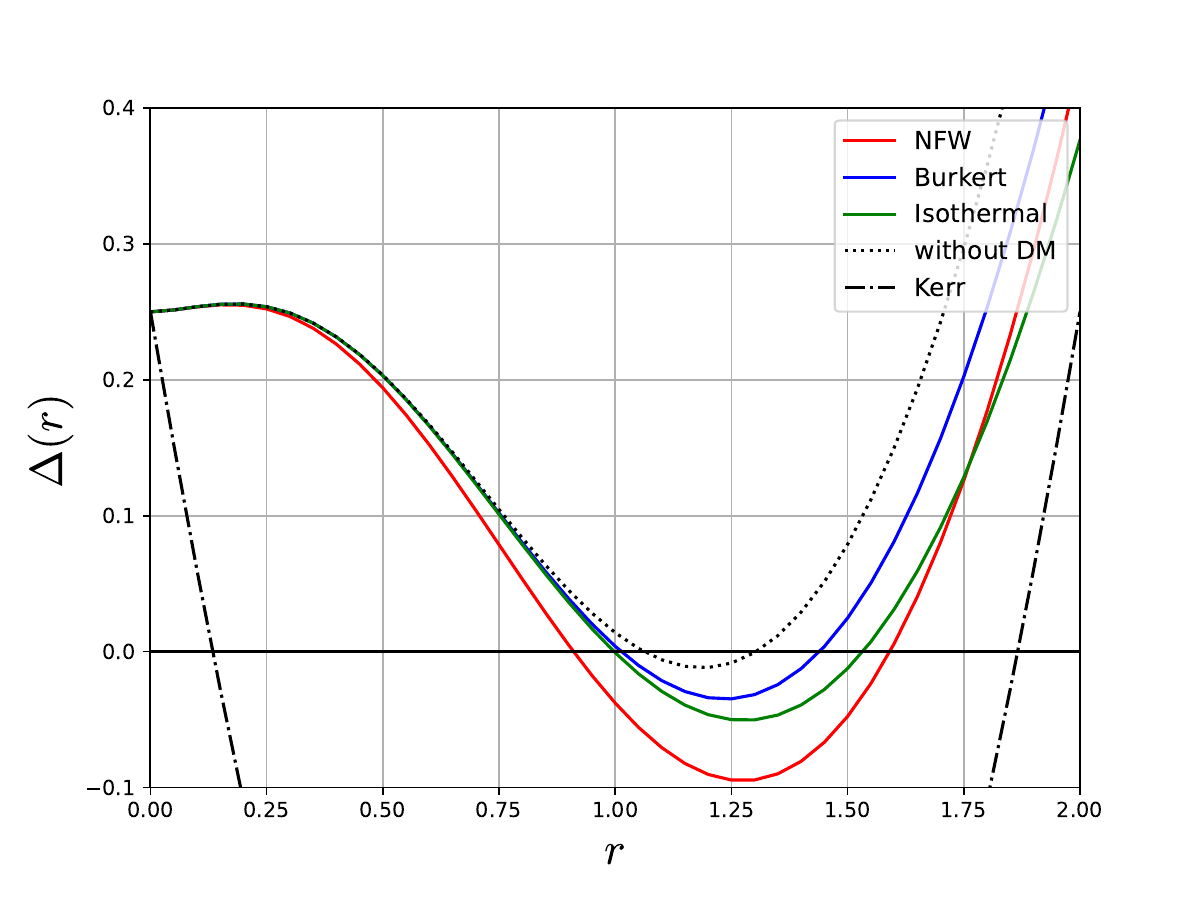}
    \caption{Behavior of $\Delta(r)$ for different DM profiles, $\rho_c=\rho_s=0.5$, $r_c=r_s=1.5$, $M=1$, $r_0=0.1$, $\alpha =0.5$ and $a = 0.5$}
    \label{figure6}
\end{figure}

The horizons associated with the solution expressed in Eq. (\ref{Romet}) can be computed via $\Delta=0$. Figure \ref{figure6} shows the position of these horizons for all dark matter profiles under consideration, as well as for the cases of EGB without dark matter ($E(r)=0$) and Kerr black hole ($E(r)=0$, $\alpha\to 0$).
Notice that Cauchy's (external) horizons are larger (smaller) than the ones of Kerr's black hole. Notice also that the pure EGB black hole has Cauchy's (external) horizons larger (smaller) than the ones of all black holes. 

\begin{figure*}[h!]
    \centering
\begin{subfigure}{.5\textwidth}
  \centering
  \includegraphics[scale=0.4]{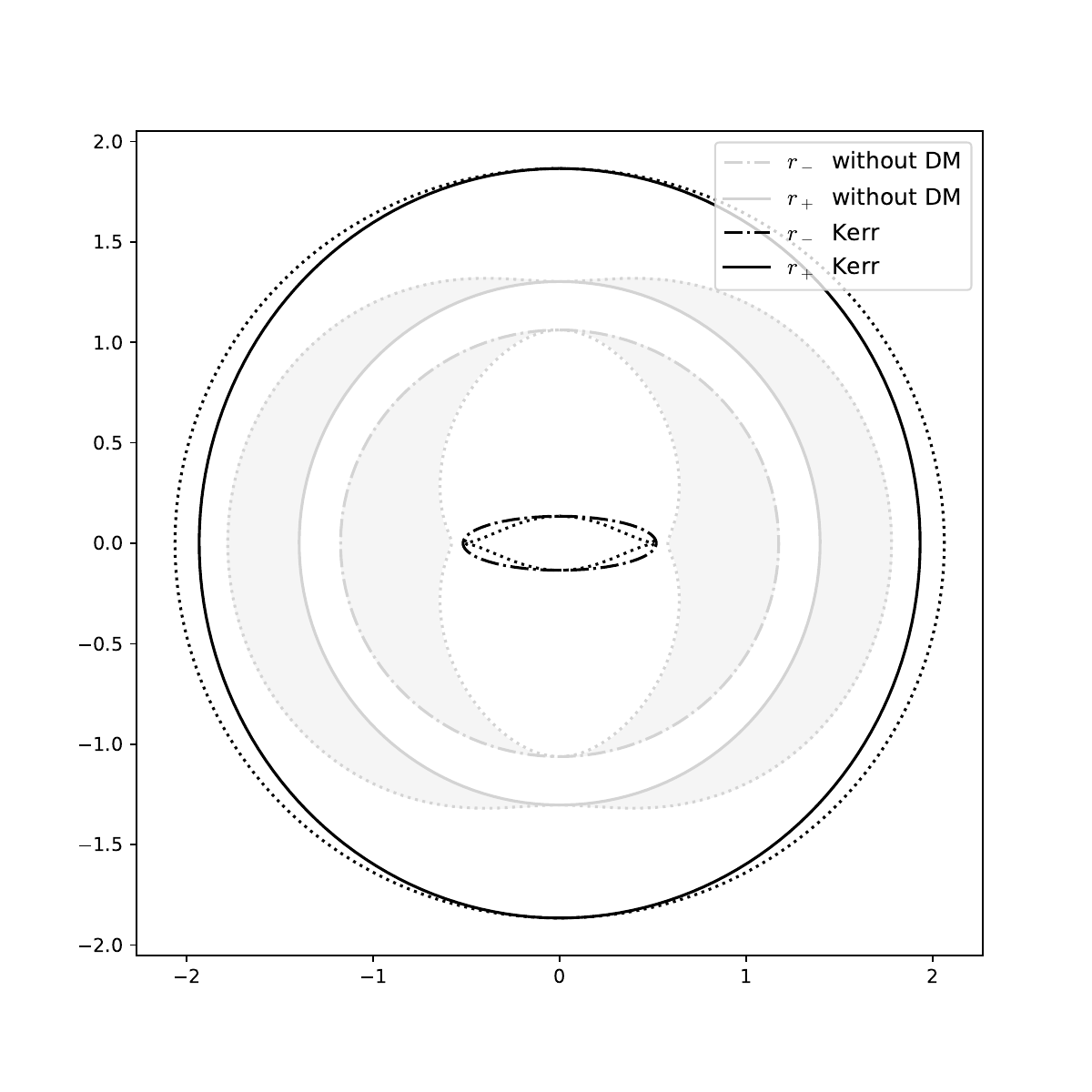}
  \caption{} 
\end{subfigure}%
\hfill
\begin{subfigure}{.5\textwidth}
  \centering
  \includegraphics[scale=0.4]{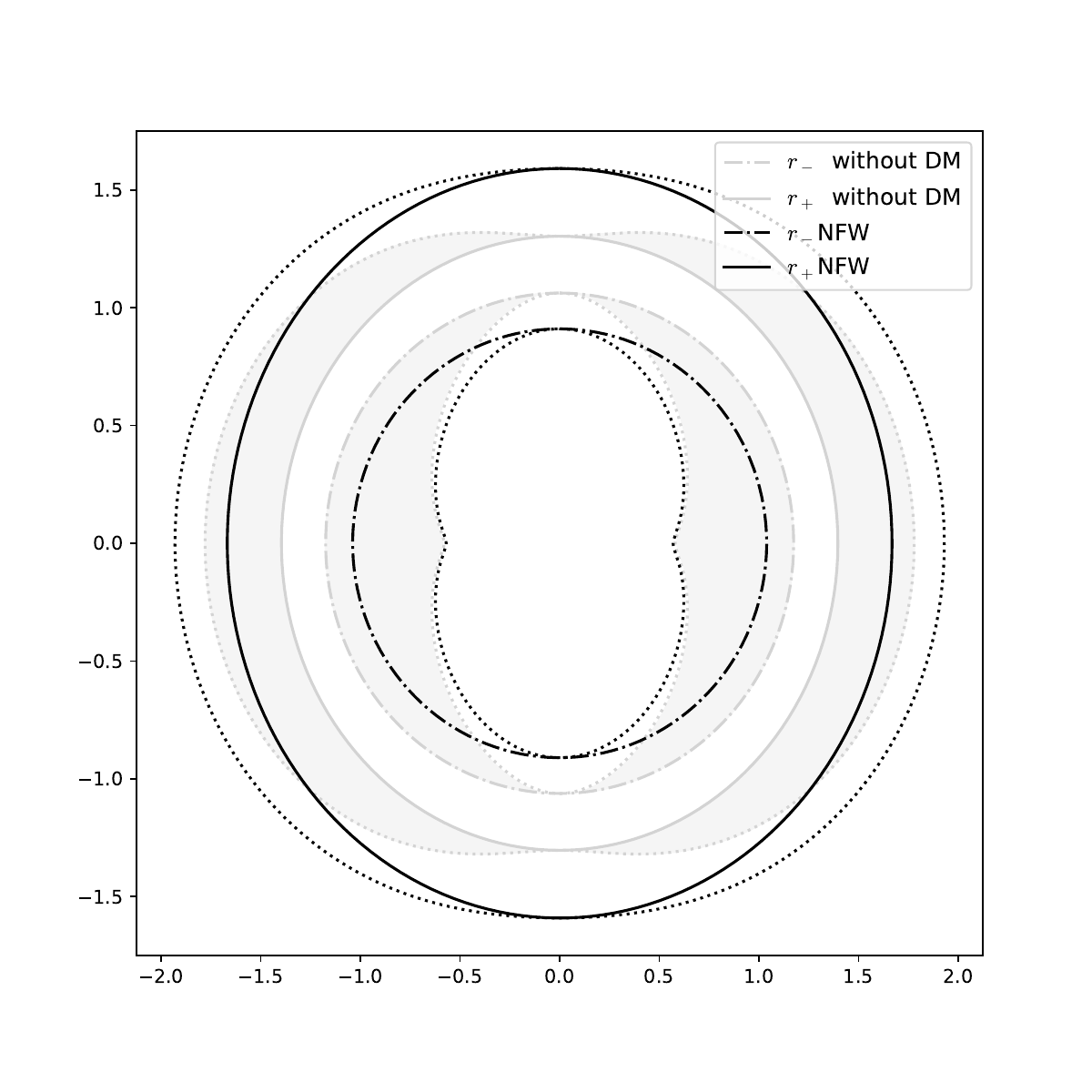}
  \caption{}
\end{subfigure}%
\\
\begin{subfigure}{.5\textwidth}
  \centering
  \includegraphics[scale=0.4]{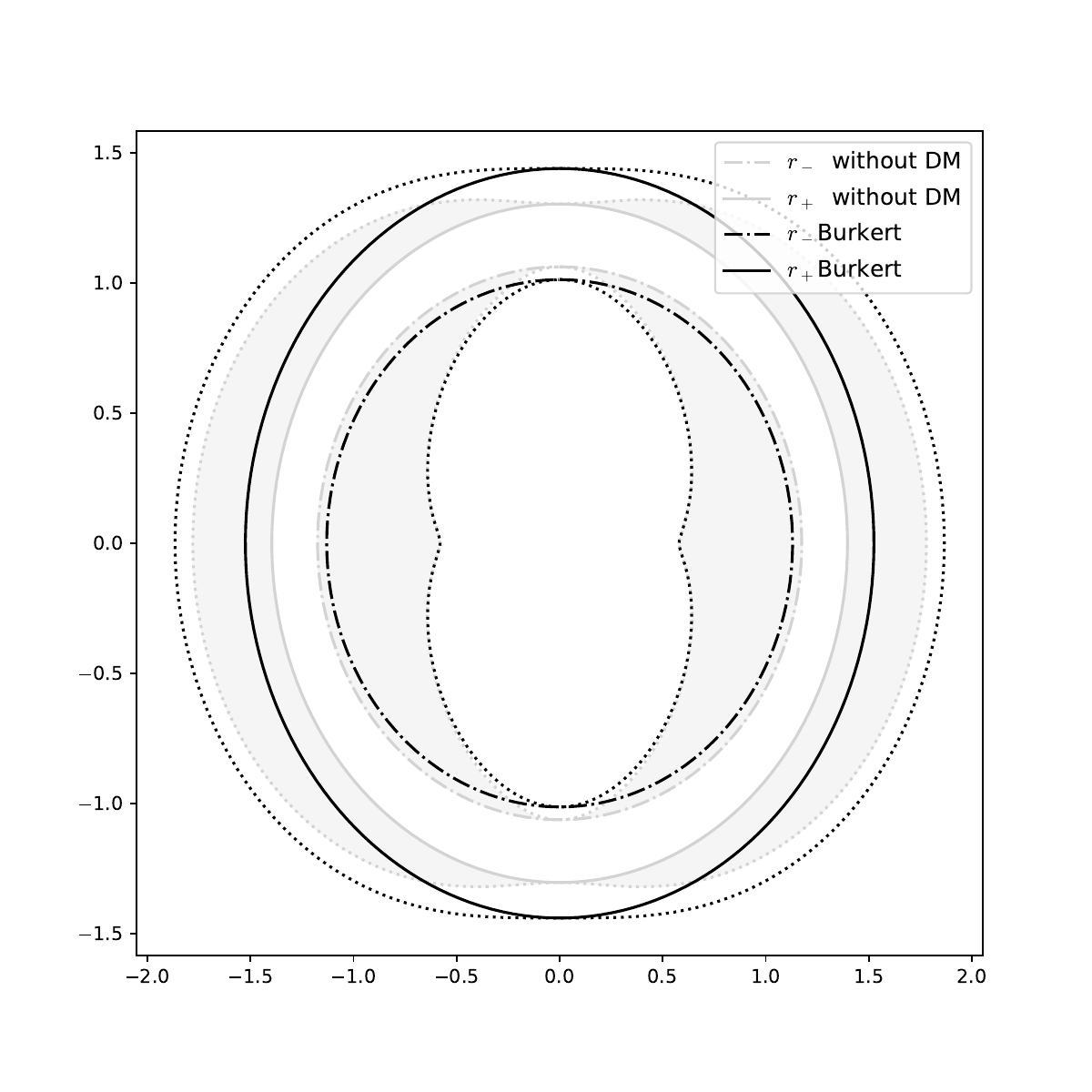}
  \caption{} 
\end{subfigure}%
\hfill
\begin{subfigure}{.5\textwidth}
  \centering
  \includegraphics[scale=0.4]{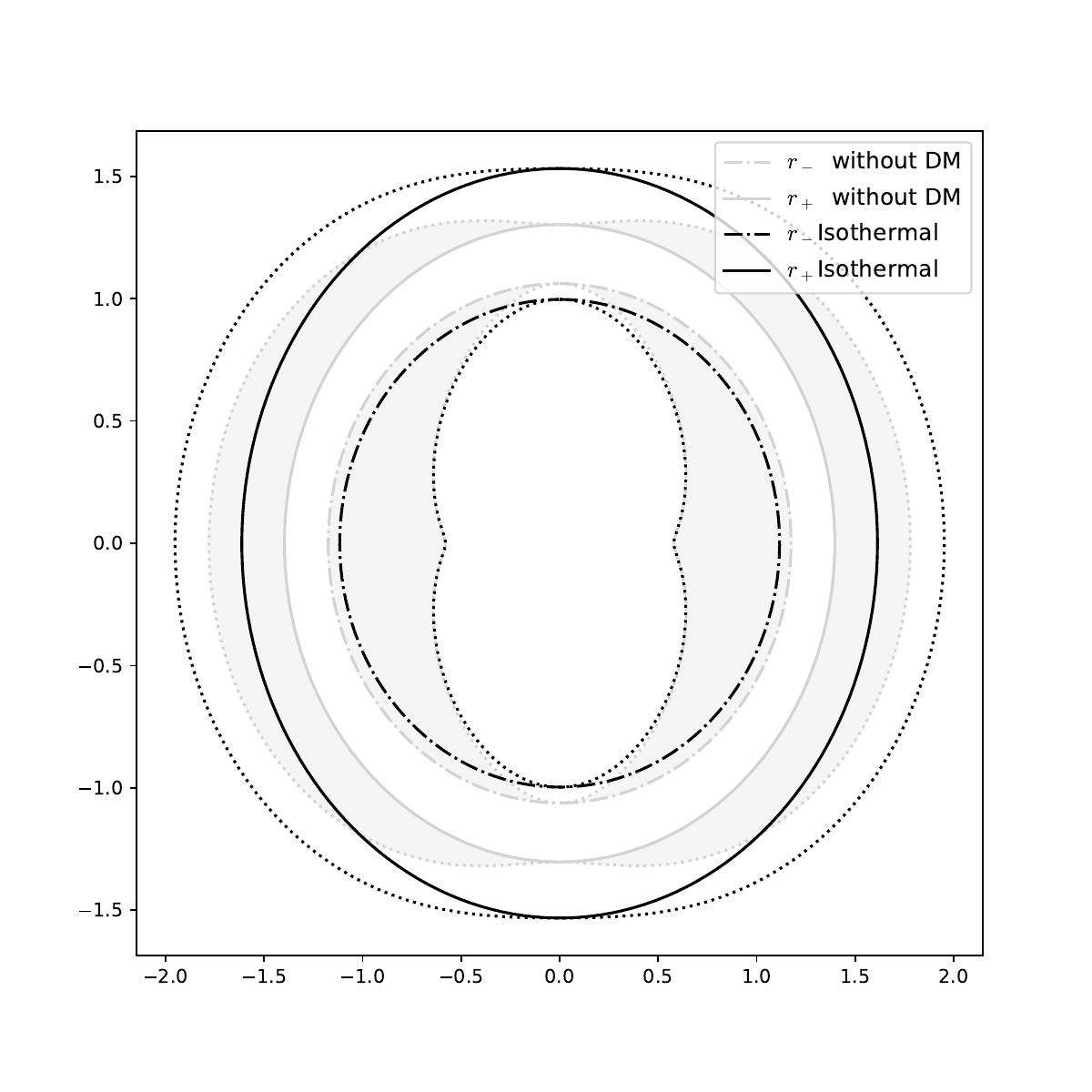}
  \caption{}
\end{subfigure}%

    \caption{The gray part of each subfigure is the same one, and it refers to EGB black hole without DM. The black part is (a) Kerr black hole, (b) EGB black hole with NFW profile, (c) EGB black hole with Burkert profile, (d) EGB black hole with Isothermal profile. In all subfigures we consider $\rho_c=\rho_s=0.5$, $r_c=r_s=1.5$, $M=1$, $\alpha =0.5$ and $a = 0.5$. }
    \label{ergo_regions}
\end{figure*}

The ergospheres are regions situated between the horizon and the static limit surface (SLS), which are given by $g_{tt}=(\Delta-a^2 \sin^2{\theta})/\Sigma=0$, where no static observer is possible \cite{Maharaj}. Thus, the external ergospheres are defined in the range $r_{+}<r<r_{+}^{SLS}$, with $r_+$ standing for the external horizons. Fig. \ref{ergo_regions} depicts the ergospheres for each black hole solution, including EGB without dark matter and Kerr black hole. Notice that Kerr's black hole ergospheres are larger than the ones of pure EGB black holes. On the other hand, the presence of dark matter makes these regions narrower with respect to those of pure EGB black holes.
\subsection{Thermodynamic}

To study the thermodynamic properties, we first obtain the Hawking temperature. For this, we need to determine the mass parameter, which is given by
\begin{equation}\label{rot_mass}
 M=\frac{r_h}{2}+\frac{\alpha}{2r_h}+\frac{a^{2}}{2r_h}+\frac{\alpha a^{2}}{r_h^{3}}+\frac{\alpha a^{4}}{2r_h^{5}}-\frac{E(r_h)}{32\pi}.
\end{equation}
Note that the above expression reduces to Eq. (\ref{16.2}) if $a=0$. The Hawking temperature is obtained by plugging the above expression into 
\begin{equation}
T_H=\frac{1}{4\pi}\frac{\Delta'(r_h)}{r_h^2+a^2},
\end{equation}
and we get
\begin{equation}\label{rot_temp}
T_{H}=\frac{1}{4\pi(r_h^{2}+a^{2})\left(r_h^{4}+2\alpha r_h^{2}+2\alpha a^{2}\right)}\left(r_h^{5}-\alpha r_h^{3}-a^{2}r_h^{3}-6\alpha a^{2}r_h-\frac{5\alpha a^{4}}{r_h}-r_h^{7}\frac{\rho(r_h)}{4}\right).
\end{equation}
Eq. (\ref{rot_temp}) provides the Hawking temperature of an EGB rotating black hole surrounded by a generic spherical density. When the rotation parameter $a$ tends to zero, the Hawking temperature in the static case is recovered, equation (\ref{18.2}). We can see in figure 2 the behavior of Hawking temperature of the EGB rotating black hole for each density dark matter profile. The stationary case also points to the existence of remnants ({\it i.e.}, when $T_H=0$), in pure EGB black holes as in those ones with the dark matter presence.
\begin{figure}[!h]
    \centering
    \includegraphics[width=10.0cm]{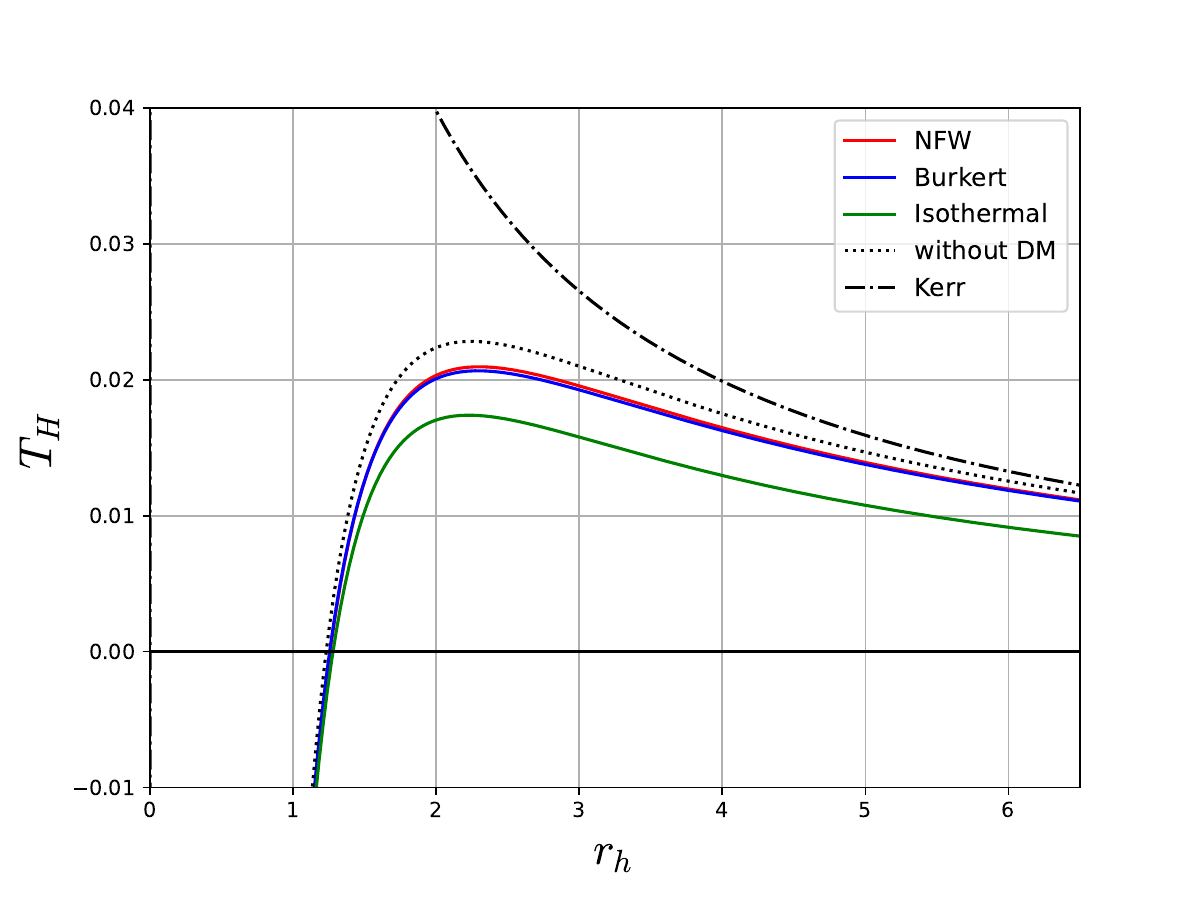}
    \caption{Behavior of $T_H$ for different DM profiles, $\rho_c=\rho_s=0.5$, $r_c=r_s=1.5$,  $\alpha =0.5$ and $a = 0.5$}
    \label{figure7}
\end{figure}

Regarding entropy, we have  
\begin{eqnarray}
    dS &=&\frac{dM}{T} -\frac{a^2}{a^2+r_h^2}\frac{dM}{T} \nonumber \\
    &=&\frac{2\pi\left(r_h^{4}+2\alpha r_h^{2}+2\alpha a^{2}\right)}{r_h^{3}}dr_h,
\end{eqnarray}
and then we can integrate it, arriving at
\begin{eqnarray}\label{rot-1}
    S = \pi (r_h^2 + a^2) -\pi (r_0^2 + a^2) + 2\pi \alpha \left[\ln\left(\frac{r_h^2}{r_0^2}\right)-\frac{a^2}{r_h^2}+\frac{a^2}{r_0^2}\right],
\end{eqnarray}
where $r_0$ is a radius associated to the integration constant of entropy. We can see in equation (\ref{rot-1}) that the entropy does not depend on the density of the matter around the black hole as in the static case. We can also see that the GB term of entropy breaks the area law $S=A/4 = 4\pi(r^2+a^2)/4$ as in the static case too. Now, besides the logarithm term, the rotation provides an additional term $2\pi\alpha a^2r_h^{-2}$.

\begin{figure}[!h]
    \centering
    \includegraphics[width=10.0cm]{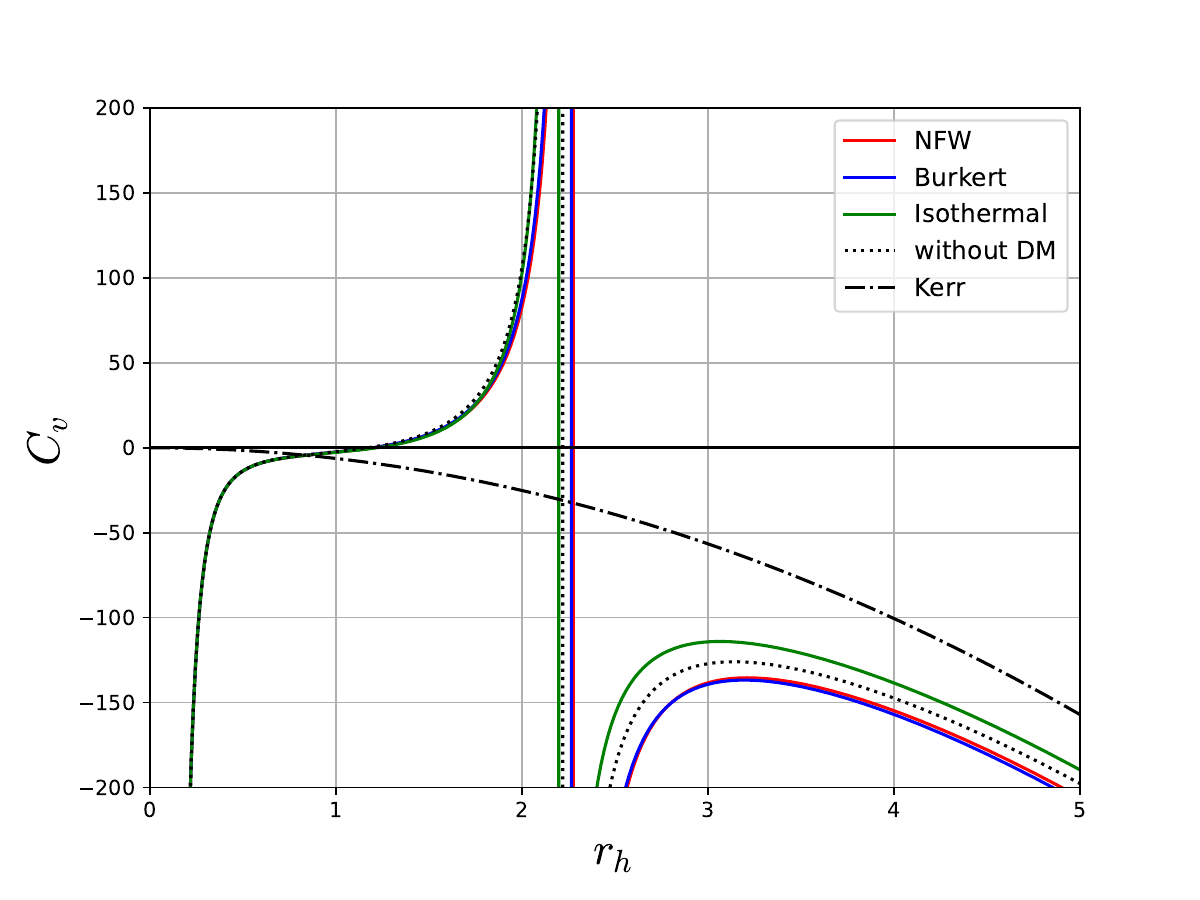}
    \caption{Behavior of $C_v$ for different DM profiles, $\rho_c=\rho_s=0.5$, $r_c=r_s=1.5$, $\alpha =0.5$ and $a = 0.5$}
    \label{figure8}
\end{figure}

The heat capacity is given by the same expression previously used, $C_v= dM/dT_H$, thus we can use equations (\ref{rot_mass}) and (\ref{rot_temp}) to compute this quantity, finding

\begin{eqnarray}\label{rot-2}
C_v &=& - \left\{\frac{2 \pi}{r_h^4} \left(a^{2} + r_h^{2}\right)^{2} \left(2 a^{2} \alpha + 2 \alpha r_h^{2} + r_h^{4}\right)^{2} \cdot \left(20 a^{4} \alpha + 24 a^{2} \alpha r_h^{2} + r_h^{6} \left(r_h^{2} \rho(r) - 4\right) + 4 r_h^{4} \left(a^{2} + \alpha\right)\right)\right\} \nonumber \\
&\times& \left\{4 r_h^{2} \left(a^{2} + r_h^{2}\right) \left(\alpha + r_h^{2}\right) \left(20 a^{4} \alpha + r_h^{2} \cdot \left(24 a^2 \alpha + r_h^{6} \rho(r_h) - 4 r_h^{4} + 4 r_h^{2} \left(a^{2} + \alpha\right)\right)\right) \right.\nonumber \\
&&\left.+ 2 r_h^{2} \left(2 a^{2} \alpha + 2 \alpha r_h^{2} + r_h^{4}\right) \cdot \left(20 a^{4} \alpha + r_h^{2} \cdot \left(24 a^2 \alpha + r_h^{6} \rho(r_h) - 4 r_h^{4} + 4 r_h^{2} \left(a^{2} + \alpha\right)\right)\right)  \right. \nonumber \\
&&\left.+ \left(a^{2} + r_h^{2}\right) \left(2 a^{2} \alpha + 2 \alpha r_h^{2} + r_h^{4}\right)\left[20 a^{4} \alpha - \left(24 a^{2} \alpha r_h^{2} + r_h^{9} \frac{d \rho}{d r_h} + 7 r_h^{8} \rho(r_h) - 20 r_h^{6} + 12 r_h^{4} \left(a^{2} + \alpha\right)\right)\right] \right\}^{-1} .\nonumber \\
\end{eqnarray}
Equation (\ref{rot-2}) tends to equation (\ref{21.2}) when the rotation parameter tends to zero. We can see in figure (\ref{figure8}) the heat capacity behavior for each density dark matter profile. As in the static case, the rotating black holes present smooth (abrupt) phase transitions from locally unstable (stable) regions to stable (unstable) ones ($C_v<0$, unstable; $C_v>0$, stable). An essential consequence of the above expression is that the numerator is null at the same $r_h$ in which the temperature is null. This enforces the conclusion about the existence of black hole remnants.

\begin{figure}[!h]
    \centering
    \includegraphics[width=10.0cm]{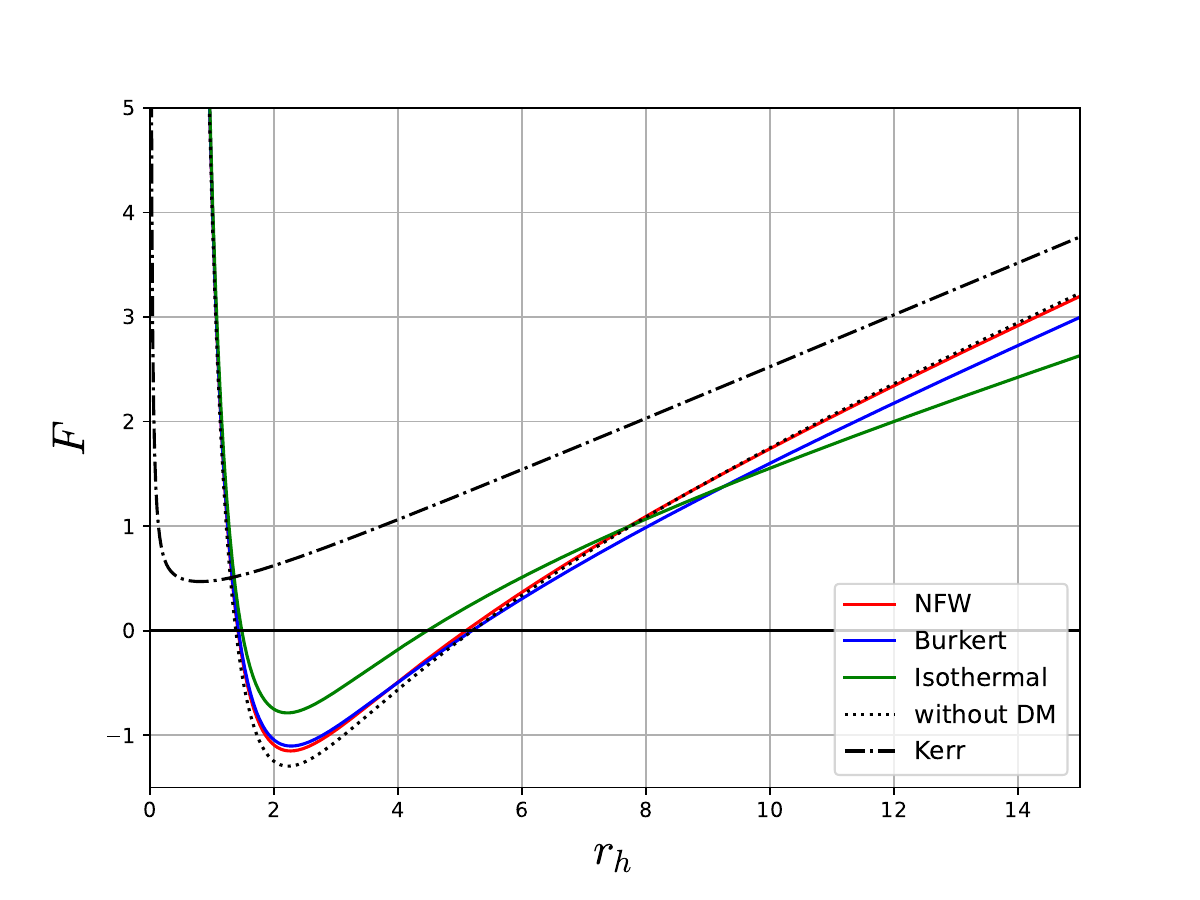}
    \caption{Behavior of $F$ for different DM profiles, $\rho_c=\rho_s=0.5$, $r_c=r_s=1.5$, $\alpha =0.5$ and $a = 0.5$}
    \label{figure9}
\end{figure}

We also compute the Gibbs free energy, $F = M - TS$, we find
\begin{eqnarray}\label{rot-3}
    F &=& \frac{r_h}{2}+\frac{\alpha}{2r_h}+\frac{a^{2}}{2r_h}+\frac{\alpha a^{2}}{r_h^{3}}+\frac{\alpha a^{4}}{2r_h^{5}}-\frac{E(r_h)}{32\pi} \nonumber \\
    &-&\left[ \frac{\left(r_h^{6}-\alpha r_h^{4}-a^{2}r_h^{4}-6\alpha a^{2}r_h^2-5\alpha a^{4}-r_h^{8}\rho(r_h)/4\right)}{4\pi r_h(r_h^{2}+a^{2})\left(r_h^{4}+2\alpha r_h^{2}+2\alpha a^{2}\right)}\right] \nonumber \\
    &\times&\left\{\pi (r_h^2 + a^2) -\pi (r_0^2 + a^2) + 2\pi \alpha \left[\ln\left(\frac{r_h^2}{r_0^2}\right)-\frac{a^2}{r_h^2}+\frac{a^2}{r_0^2}\right]\right\}. 
\end{eqnarray}
Equation (\ref{rot-3}) tends to equation (\ref{22.2}) when the rotation parameter tends to zero. We can see in figure (\ref{figure9}) the behavior of free energy for each density dark matter profile. We can also see in figure (\ref{figure9}) that there are only smooth (global) phase transitions ($F<0$, stable; $F>0$, unstable).
\section{Conclusion}\label{section-6}

In this paper, we have obtained exact black hole solutions surrounded by dark matter halos in Einstein--Gauss--Bonnet gravity and studied their thermodynamic properties. Thus, in section 2, we found the static EGB black hole solution with a generic spherical distribution (density) of matter. In section \ref{section-2},  we have analytically found the corresponding thermodynamic properties: temperature, entropy,  constant-volume heat capacity, and Gibbs free energy (see equations (\ref{18.2}), (\ref{20.2}), (\ref{21.2}), (\ref{22.2})).

In section \ref{section-3}, we analyzed the EGB black hole solution particularizing to the following phenomenological density profiles of dark matter: (i) Burkert, (ii) NFW, and (iii) pseudo-isothermal. The behavior of the solution summarized in equation (\ref{7}) depends on the EGB coupling constant $\alpha$, as well as on the dark matter characteristic parameters of density and radius. This solution can present zero, one, or two horizons according to the setting of parameters. Figure \ref{figure1} exhibits the solution behavior with two horizons for each dark matter profile. It is worth pointing out that these horizons are smaller than the one associated with the Schwarzschild black hole solution.

Using the same set of parameters, we studied the thermodynamic properties. Figure \ref{figure3} presents the temperature for each dark matter profile according to equation (\ref{18.2}). The interesting point is that the temperature goes to zero for every horizon radius which satisfies equation (\ref{16}), indicating the emergence of a remnant when the black hole halts its evaporation. Figure \ref{figure3} graphically shows these points for all dark matter cases. We also saw that a generic spherically symmetric distribution of matter as a gravitational source does not contribute to the entropy in the equation (\ref{20.2}). Then the entropy is the same as a pure EGB black hole. 

We also analyzed the constant-volume heat capacity and the Gibbs free energy. We presented the analytical expression of the former, for each dark matter profile,  in equation (\ref{21.2}). We analyzed its behavior in figure \ref{figure4} and identified continuous and discontinuous (local) phase transitions for critical horizon radii. We also analyzed the behavior of Gibbs free energy for each dark matter profile in figure \ref{figure5}, according to equation (\ref{22.2}). In terms of this quantity, the EGB black hole surrounded by dark matter exhibits continuous (global) phase transitions.  

We studied the more relevant rotating case in section \ref{rotating}. We split it into two parts. In the first one, we used the Newman-Janis algorithm to find the metric of a rotating 4D EGB black hole surrounded by a spherical density of matter (see equations (\ref{Romet}), (\ref{SIGMA}), (\ref{DDm})). As we did to static black hole solution,  we have analytically found the thermodynamic properties of rotating solution: temperature, entropy, constant-volume heat capacity, and Gibbs free energy (see equations (\ref{rot_temp}), (\ref{rot-1}), (\ref{rot-2}), (\ref{rot-3})). We also studied the influence of each phenomenological DM profile on the solution and thermodynamic properties. Figure \ref{figure6} shows the behavior of  $\Delta(r)$ for each dark matter profile, and figure \ref{ergo_regions} shows the inner (Cauchy) and external radii horizon and the ergosurfaces of Kerr and DM profile solutions compared to the EGB rotation solution without dark matter.

 Next, we analyzed the thermodynamic properties of the rotating case. Figure \ref{figure7} presents the temperature according to each dark matter profile. As occurs in the static case, the temperature goes to zero for every horizon radius which satisfies 
$$
4r_h^{4}(r_h^{2}-\alpha -a^{2})-24\alpha a^{2}r_h^2-20\alpha a^{4}-r_h^{8}\rho(r_h) = 0,
$$
indicating the emergence of a remnant when the rotating black hole halts its evaporation, as in the static case. We can focus on DM distributions when we consider the specific density profiles. We also saw that a generic spherically symmetric distribution of matter does not change the entropy (equation (\ref{rot-2})), as also in the static case. 
We extend our analysis of constant-volume heat capacity and Gibbs free energy to the stationary solution. Figure \ref{figure8} shows the behavior of the heat capacity of the rotating solution for each dark matter profile. We identified continuous and discontinuous phase transitions like in the static case. Gibbs free energy associated with the rotating solution with DM presented in figure \ref{figure9} exhibits continuous (global) phase transitions like in the static case.

An important feature shared by the static and rotating cases is that for both the heat capacity is zero always that the temperature is zero. This enforces a phase transition at this value of the horizon radius. Finally, as far as we know,  this is the first analytical solution to the metric in the EGB gravity in the presence of DM. Beyond thermodynamical quantities,  it allows the study of shadows, ISCO, and many other features of galactic centers. Some of these are the subject of research by the present authors. 

\section*{Acknowledgement}
The authors would like to thanks Alexandra Elbakyan and sci-hub, for removing
all barriers in the way of science We acknowledge the financial support provided by the Conselho Nacional de Desenvolvimento Científico e Tecnológico (CNPq), the Coordenação de Aperfeiçoamento de Pessoal de Nível Superior (CAPES) and Fundaçao Cearense de Apoio ao Desenvolvimento Científico e
Tecnológico (FUNCAP) through PRONEM PNE0112- 00085.01.00/16.

\end{document}